%% file: 00-manuscript.tex
\documentclass[journal]{vgtc}                     

\onlineid{0}

\vgtccategory{Research}

\vgtcpapertype{Theoretical and Empirical}

\title{Evaluating Affective Objectives: Statistical Numbing in Data Visualization}

\author{%
  \authororcid{Elsie Lee-Robbins}{0000-0002-4080-6506}
    and 
  \authororcid{Eytan Adar}{0000-0003-1911-836X}
}

\authorfooter{
  \item
  Eytan Adar is with the University of Michigan. Elsie Lee-Robbins conducted this research during her PhD program at the University of Michigan. E-mail: \{elsielee\,$|$\,eadar\}@umich.edu\,
}

\abstract{%
  Visualizations can help audiences understand the scale of tragedies, such as the consequences of natural disasters, war, genocide, and pandemics. In these cases, a visualization designer's default behavior may be to focus on communicating quantitative information: numbers, statistics, and trends. However, this may not reflect higher-level affective objectives to inspire their audience to care about an issue, empathize with others, or take action to help those in need. Worse, standard visualizations may conflict with these goals, as statistics can numb emotions and reduce prosocial feelings toward people in need. Designers have developed strategies to increase affective responses through data visualizations, such as blending data narratives and personal narratives about individuals. In this paper, we explore three design strategies for communicating a humanitarian crisis: data-driven, human-driven, or mixed narratives. We conducted an empirical study to explore the effect of statistical numbing in the context of these types of narratives in the format of data videos. In particular, we measure prosocial feelings and behaviors by giving participants the option of donating money as part of the study. We find that human-driven narratives (photographs and stories of individuals) elicited the highest donations and that the mixed narrative combination led to the lowest donations. We discuss the limitations of this study and the implications of pursuing affective objectives and the numbing of empathy in data visualization design.
}


\teaser{
  \centering
  \includegraphics[width=\linewidth, alt={}]{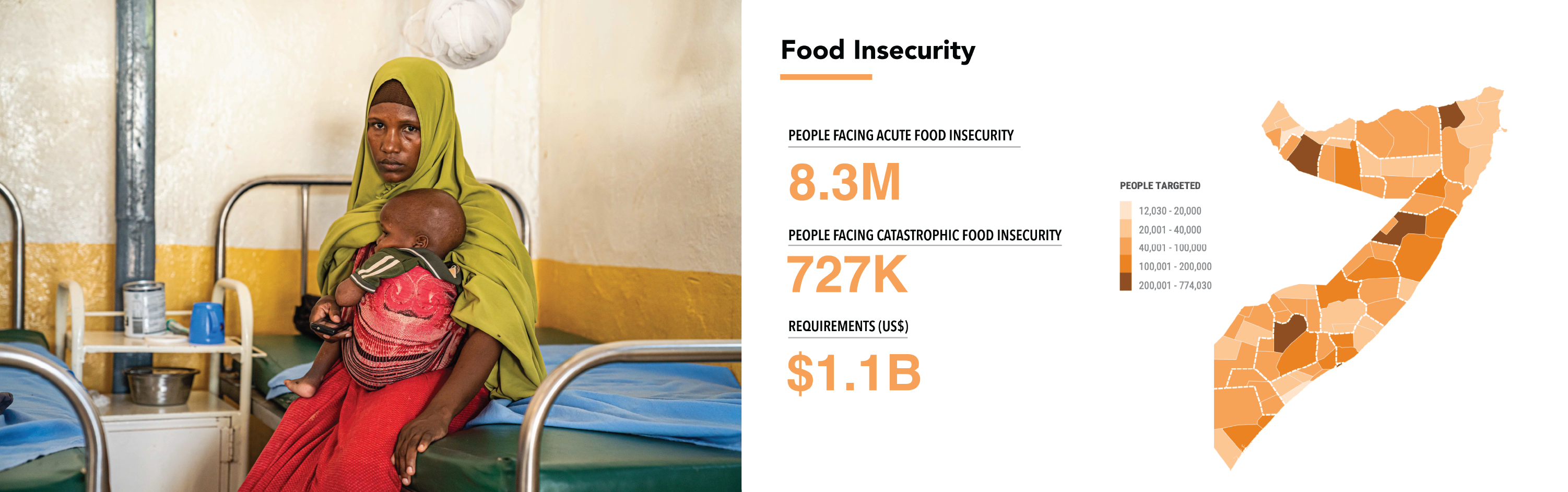}
  \caption{%
  	(Left) A picture of Ruqio, a Somali woman, and her child in a hospital. Image from the United Nations Office for the Coordination of Humanitarian Affairs (UN OCHA) \cite{UNOCHA_2022}. (Right) Data visualization of the population and location of Somalis facing food insecurity. Chart recreated, based on UN OCHA's Humanitarian Response Plan Report \cite{UNOCHA_2023}. Which is more effective to inspire prosocial feelings and behaviors?%
  }
  \label{fig:teaser}
}

\renewcommand{\manuscriptnotetxt}{}

\nocopyrightspace

\graphicspath{{figs/}{figures/}{pictures/}{images/}{./}} 

\usepackage{tabu}                      
\usepackage{booktabs}                  
\usepackage{lipsum}                    
\usepackage{mwe}                       

\usepackage{mathptmx}                  
\usepackage{balance}
\usepackage[hyphens]{url}

\begin{document}



\maketitle

\input{01-introduction.tex}

\input{02-relatedwork.tex}

\input{03-study.tex}

\input{04-discussion.tex}

\input{05-conclusion.tex}




\acknowledgments{%
    Thank you to Peiying Loh for sharing her thoughts on Kontinentalist's work and to Alison Siegler, Francis Gagnon, Gabrielle Merite, Dino Citraro, and Peiying Loh for their permission to reuse their images. The authors thank the NSF for their support of this work through NSF IIS-1815760. This work was supported in part by funding from the Rackham Graduate School at the University of Michigan.%
}

\balance

\bibliographystyle{abbrv-doi-hyperref}

\bibliography{bib}

\end{document}

%% file: 01-introduction.tex
\section{Introduction} 

Advocacy groups and organizations with a mission to improve the world often use data collected about humanitarian crises. Their goals are to raise awareness of the crisis and elicit responses from their audience: donate money to the cause, share the message with friends and family, or contact policymakers. Unlike many visualizations, the focus is not on achieving \textit{cognitive} goals (recall facts, analyze trends, evaluate the data, etc.)~\cite{adar2021}. Instead, the goals are \textit{affective}: they aim to influence the audience's appraisals, attitudes, or values~\cite{lee2022affective}.  In some ways, designing communicative visualizations for cognitive goals is more straightforward. There is significant guidance on their construction, significant research on their effectiveness, and there are rarely alternatives to visualizations. Cognitive goals are so entangled with the data itself, that it is hard to argue that visualizations are not nearly always optimal for this task. With affective goals, there is far less guidance and research. More critically, there are other alternatives to data visualizations to achieve affective goals. Photography is often the most popular choice for certain types of advocacy communication~\cite{rall2016data} but other types of illustration and visual art are also common.

In their efforts to achieve affective goals, organizations and designers use a variety of strategies. One mechanism is to use an emotional appeal (the rhetorical pathos) to increase empathy for suffering people. Tactics for an emotional appeal could be photographs, audio, videos, and stories about individuals. Photojournalism, as a journalistic form, brings the story to life and gives people a deeper understanding of the story. Using an emotional appeal might result in a narrative that is focused on individuals who are in need.

In addition to an emotional appeal, organizations may also want to make a logical argument (the rhetorical logos). This strategy often involves the use of statistics and numbers to draw a more compelling picture of an issue by presenting data as evidence. Numbers and statistics can be presented as facts, wielded as reasons to support a persuasive argument. Data visualization is one tool that can effectively show trends, patterns, and the scope of problems. Showing the large scale of a humanitarian crisis would be a compelling argument for why it is important; a humanitarian crisis that impacts millions of people could grab attention. Although we believe that data may work for affective goals, humans do not naturally respond to statistics with empathy~\cite{Slovic_Slovic_2015}. Exclusively logical arguments may not work as well as emotional ones or some combination of strategies. 

Some organizations try to combine both strategies by integrating data-driven- with human-driven-narratives as a best-of-both-worlds strategy (see Figure \ref{fig:examples} for examples). They hope to create a richer and more affective experience for the viewer. Some designers consider just the data insufficient and dehumanizing, and naturally want to include stories from real people to bring context to the data (Peiying Loh, personal communication, May 28, 2024). Data visualization research has studied the potential for visualizations to increase the emotional response of an audience. Much of this research focuses on \textit{anthropographics}, which are visualizations about people that are designed to evoke empathy and promote prosocial behaviors. An example of a data visualization-specific design element is to increase granularity so that each mark represents one person~\cite{Morais_Jansen_Andrade_Dragicevic_2020}. This design decision would look like disaggregating bars or squares to show each person as their own individual mark. In this way, the data visualization designer introduces some of the human element back into the statistics. 
Many of these studies focus on the comparison of anthropographics with traditional visualization designs, such as bar charts and pie charts~\cite{Boy_Pandey_Emerson_2017, Campbell_Offenhuber_2019, Morais_Dandara_Sousa_Andrade_2020, Morais_Jansen_Andrade_Dragicevic_2021, Liem_Perin_Wood_2020}. However, these studies have not found much success with these designs. Evidence shows that there is only a very small effect on prosocial behaviors (for a review, see~\cite{Morais_Jansen_Andrade_Dragicevic_2021}). 

There are other ways to integrate stories and photographs into the data without creating an anthropographic. A hybrid design could present information about individuals separately from the visualizations themselves. This allows for a few advantages. First, it allows the designers to use datasets that do not have rich, detailed information about the people in the data. 
By separating the data set from the stories and photographs of individuals, the designer can present rich information about individuals that might be hard to visually present at the same time as a visualization. These two types of information can be presented sequentially. For example, a data story about Rohingya refugees combines a narrative with illustrations of refugees, an audio snippet from a real refugee, and maps that show their perilous journey at sea~\cite{Kontinentalist_2020}. Each of these segments are viewed sequentially, through an interactive `scrollytelling' audio-visual narrative. 

However, most data-oriented designers do not empirically evaluate their intuitions: that data and visualizations help achieve affective objectives. They do not test whether their designs are more effective with the combination of data visualizations and stories of individuals compared to either option alone. 
The reality is that psychology research suggests that data visualizations are less effective in increasing empathy and prosocial behaviors. Statistical numbing is an effect where showing statistics about the broader issue decreases empathy~\cite{Small_Loewenstein_Slovic_2007}, while empathy is highest for a single individual in need~\cite{Kogut_Ritov_2005, Vastfjall_Slovic_Mayorga_Peters_2014}. 
Therefore, if a designer's goal is to elicit empathy and encourage action, data visualizations could potentially have an \textit{adverse effect} on eliciting empathy. Data visualization \textit{itself} might be the wrong solution. Data visualization brings attention to the full scope of the problem, showcasing many, many people, perhaps in abstraction. Showing a photograph and a story about one individual can be a more effective way to evoke empathy and prosocial behaviors. That is, maybe it is better not to show a data visualization at all.

Here, we apply the framework of affective learning objectives~\cite{lee2022affective} to evaluate the effectiveness of three designs. Instead of starting with the data visualization design, we start with describing our goals---in this case, affective goals to elicit prosocial feelings and behaviors. Then, we consider multiple techniques on how we could achieve our goals. We build on prior work to compare data-driven narratives and an mixed narrative that combines data narratives with human narratives. However, we also compare these two designs with a third: just the human-driven narratives. Our work also contributes a novel design that utilizes a combination of self-reports and actual behaviors (donation). We provide some evidence that the \textit{combination} of both types of narratives is not necessarily the ``best of both worlds.'' In our case, we found that the human-driven narrative was the most effective at achieving our affective goals. Surprisingly, the mixed narrative performed the worst. This has implications to challenge beliefs about how to achieve affective goals with data visualizations.

We note that our analysis does not disqualify the use of data visualizations to achieve affective goals. While this paper explores one design on how to combine visualizations and rich, impactful information about individuals, there are multiple other ways to integrate this information, as well as novel ways that will be creatively imagined in the future.  The design space of all possible affective visualization designs is too large to fully cover~\cite{Morais_Jansen_Andrade_Dragicevic_2020, lan2023affective}. Rather, our work demonstrates that (a) there are situations where data visualizations can run afoul of achieving affective objectives, and (b) specifically framing the task through a learning objectives framework allows designers to validate their designs. The affective objectives framework can be a useful tool, providing structure for defining goals, and then generating and implementing evaluations. Through our study, we provide one example of how to articulate affective goals and create and deploy evaluations. Although the scale of our study is unlikely to be replicated by the average designer, the idea of creating affective evaluations is something that anyone can implement on a small scale.
Under this framework, designers can test their intuitions and empirically evaluate affective designs. They can use the results to determine which designs to choose and how (or if) to integrate data visualizations, stories, and photographs of individuals.

%% file: 02-relatedwork.tex
\section{Related Work}
\label{section:related}

\subsection{Affective Visualizations}

Affective visualizations have gained more attention as researchers consider noncognitive intents~\cite{lan2023affective,lee2022affective}. The term ``affective'' has several definitions in this community. However, as we defined in earlier work, we view affective goals as those that aim to influence appraisals, attitudes, or values. Affective visualizations are visualizations that are created in order to achieve one of these goals. 
These affective visualizations often fall under the category of anthropographics, which have been defined as ``\textit{visualizations that represent data about people in a way intended to promote prosocial feelings (e.g., compassion or empathy) or prosocial behavior (e.g., donating or helping)}''~\cite{Morais_Jansen_Andrade_Dragicevic_2020}. 

In this section, we focus on the use of affective visualizations, particularly their role in advocacy. To motivate the use of our stimuli, we discuss common affective design decisions and review a few examples of visualizations that combine data visualizations with photographs/stories of individuals. 
Finally, we review research on prosocial feelings and behaviors and their intersection with evaluating data visualization.

\subsubsection{Visualizing Data for Advocacy}

At their core, advocacy groups aim to help other people. These affective goals can be achieved in various ways. Sometimes, they launch advertising campaigns to raise awareness about an issue. Often, they organize fundraising drives to get donations for their cause. 
To achieve these goals, advocacy groups can present different types of information to their audience. One common strategy is to present testimonials, stories and in-depth information about individuals. For example, a human rights organization may publish a long-form article with vignettes about refugees. Typically, a photograph of an individual can be featured in a fundraising campaign. A motivation for this is that stories and photographs of people capture attention and can evoke an emotional connection (a \textit{pathos} rhetorical strategy)~\cite{bogre2012photography}. 

Advocacy groups can also use data visualization to make a logical appeal to their audience~\cite{Tactile_Technology_Collective_2013, ganesh2015communicating}. Nonprofit and government organizations have increasingly used data visualizations in advocacy reports in recent decades~\cite{rall2016data}. There are several reasons why advocacy groups might want to present data visualizations. First, data could give advocacy groups credibility and trustworthiness~\cite{rall2016data} (an \textit{ethos} rhetorical strategy). Data provides a logical argument and `evidence' to back up claims. Second, some designers perceive data visualizations to be persuasive~\cite{Pandey_Manivannan_Nov_Satterthwaite_Bertini_2014, rall2016data} (a \textit{logos} strategy). Third, data visualizations could simplify and distill complex information and make it accessible to a wider audience~\cite{ganesh2015communicating}.

\subsubsection{Affective Designs}

\begin{figure*}[t]
    \centering
    \includegraphics[width=\textwidth]{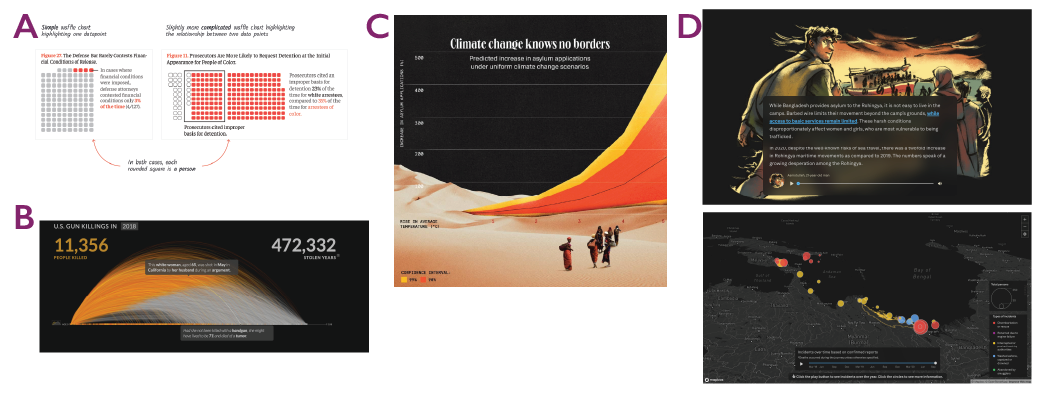}
    \caption{Examples of several affective visualizations. (A) Freeing the US from its culture of detention: waffle chart, from Voila \cite{Voila2023}. (B) US Gun Deaths, from Periscopic \cite{Periscopic_2013}. (C) Climate change knows no borders, from Gabrielle Merite \cite{Merite_2021}. (D) Two screenshots of Abandoned at Sea, from Kontinentalist in partnership with UN HCR \cite{Kontinentalist_2020}. The top screenshot shows an illustration of a refugee walking towards a boat, part of a narrative, and an audio file from Asmotulleh, a refugee. The bottom screenshot shows a map of refugees abandoned at sea. 
    }
    \label{fig:examples}
\end{figure*}

Despite all the benefits of using data visualization for advocacy, it can be perceived as cold and dehumanizing. 
Therefore, designers have explored various ways to creatively design visualizations to incorporate more of a human element and to evoke more emotions as a route to achieving affective goals.

Data visualization designers often want to humanize the data when showing sensitive topics. For example, one report on the culture of detention in the US represented the people in the federal justice system who were unjustly jailed~\cite{siegler2022freedom}. The designers of the visualizations aimed to humanize the data: ``\textit{the visuals in the report needed to make each of those humans visible, rather than `aggregating' them in one big column or bar\ldots keeping the people affected at the center of the narrative}''~\cite{Voila2023}. They described their design decision of granularity to represent each person as a separate mark (see Figure \ref{fig:examples}a). However, there are other ways to design visualizations to add humanness back into the data. Another example of an affective visualization is ``US Gun Deaths,'' which also uses granularity to show each person as their own mark~\cite{Periscopic_2013} (see Figure \ref{fig:examples}b). They also include information about each person, listing details about how they were killed and their stolen years. This design adds specificity to each mark with distinct attributes, making each distinguishable.

An anthropographic design space outlines various dimensions of how and what to visualize when visualizing data about people~\cite{Boy_Pandey_Emerson_2017, Morais_Jansen_Andrade_Dragicevic_2020}. Past work has identified seven dimensions of anthropomorphic design: granularity, specificity, coverage, authenticity, realism, physicality, and situatedness. 
Various data visualization designs in this space have been investigated to see their effect on prosocial feelings and behaviors~\cite{Boy_Pandey_Emerson_2017, Campbell_Offenhuber_2019, Morais_Dandara_Sousa_Andrade_2020, Morais_Jansen_Andrade_Dragicevic_2021, Liem_Perin_Wood_2020}. However, these designs have been suggested to only have a negligible effect~\cite{Morais_Jansen_Andrade_Dragicevic_2021}. 

In addition, other affective techniques could be used to increase an emotional response in the viewer, such as personalization, sensation, and narratives~\cite{campbell_2018, lan2023affective, kostelnick2016}. The US Gun Deaths visualization not only has elements of anthropographics, but they also use animation, pacing, and a somber style. They visualize a calculated variable called `years stolen' which highlights the loss of gun violence. In addition, designers can increase an affective response through the context in which the data visualization is seen. For example, `Climate change knows no borders' incorporates photographs of refugees, but they are not strictly tied to the data in the visualization (see Figure \ref{fig:examples}c)~\cite{Merite_2021}. Conceptually, the people depicted are more similar to ``visual embellishments'' as they are not part of the data set, but add to the design of the visualization. Other ways that designers could increase the overall affective response of the viewer is to set the visualizations within a larger narrative. One example of this is ``Abandoned at Sea'' which is a scrollytelling data story~\cite{Kontinentalist_2020}. This story combines several different elements: illustrations, audio files of specific and real refugees telling their story, maps, and a Sankey diagram. Depending on the context, visual elements such as illustrations could be less effective at achieving communicative goals~\cite{garreton2023attitudinal}. Each of these elements adds to the overall experience of the viewer to get a better understanding and a deeper sense of empathy for the refugees.

\textbf{Show each individual}---One humanizing strategy specific to data visualization is to show each individual as a mark~\cite{Rost2017}. The visualization dimension of granularity is how much the chart shows the data aggregated or individually~\cite{Morais_Jansen_Andrade_Dragicevic_2020}. A low-granularity data visualization could be a bar chart with the bars representing an aggregated statistic about the dataset. On the other hand, designers can represent each individual as their own mark, whether that is a square (``Out of Sight, Out of Mind'': Drone strikes in Pakistan by Pitch interactive~\cite{Grubbs_Madrid_Yahnke_Lin_2013}), line (``U.S. Gun Deaths'' by Periscopic~\cite{Periscopic_2013}), icon (Black by the Numbers~\cite{Black_2019}), or their name (New York Times~\cite{NYT_names_2020}). This design strategy can more intensely portray the scope of the dataset and remind the audience that each of the marks is a real and unique person.

\textbf{Put a face to the data}---A strategy to highlight the humans in the data is to show their faces. Photojournalism takes advantage of this by often portraying humans in their work~\cite{Lewis_2020, Zeeberg_2016}. Faces are a reminder that the dataset is made up of real humans; designers should ``show what the data are about''~\cite{Rost2017}. People are evolved to recognize and perceive faces, so they are a powerful and realistic way to represent people, compared to icons of people~\cite{Morais_Jansen_Andrade_Dragicevic_2020}. Some argue that this is a reason not to use data visualization, as pictures of faces can be more effective than visualized data~\cite{Harris2015}. In trying to design a visualization about children with fatal diseases, one designer realized that the best solution was not a visualization at all --- it was simply pictures of the children~\cite{Slobin}.

\textbf{Emphasizing specific individuals}---The journalist Nicholas Kristof uses the power of focusing on individuals to engage his readers~\cite{Kristof_2009}. This strategy is powerful because it is easier to form a mental image of and relate to a single person compared to a group of people. One example of this is the photograph of Alan Kurdi, which sparked empathy for Syrian refugees that had not been elicited in previous news coverage~\cite{Demir_2017}. 
Harris suggests the principle ``Near and Far'' to focus on a smaller range of data along with the larger picture~\cite{Harris2015}. Similarly, Rost suggests to ``Zoom into one dot'' and highlight the details of one person~\cite{Rost2017}. Additionally, specificity is a dimension of data visualization that could humanize the data by incorporating specific information about the people portrayed~\cite{Morais_Jansen_Andrade_Dragicevic_2020}. However, it is not always possible or desired to have full details of every person in the dataset. In sensitive data cases, designers may withhold identifiable information for privacy or security~\cite{Boy_Pandey_Emerson_2017}. 
In addition, it can sometimes backfire to focus on individuals. In cases where public opinion ascribes personal blame (e.g., obesity), showing individuals can lead to less support than a more general framing of the problem~\cite{barry2013framing}.

\textbf{Data Storytelling}---Storytelling speaks to the reader's imagination and the words of the story paint a vivid picture. The author can use descriptive and figurative language to evoke emotions. In journalism, a typical news story can be elevated by adding personal narratives, mobilizing information, and photography~\cite{Maier_Slovic_Mayorga_2017}. 
Storytelling can engage readers, evoke emotions, and take advantage of curiosity and surprise to keep them interested~\cite{kostelnick2016}. Within storytelling, there are many rhetorical techniques that designers can use to frame their story and guide the reader~\cite{Hullman_Diakopoulos_2011}. Data comics could be another way to provide more of a structured personal narrative with visualizations, even potentially eliciting more emotions with this format~\cite{Bach_Wang_Farinella_Murray-Rust_Riche_2018, Alamalhodaei_Alberda_Feigenbaum_2020, Emerson_Satterthwaite_Pandey_2018}.

\subsection{Statistical Numbing}

Although there is a desire for organizations to visualize their data, they often make choices on when and how to visualize data from anecdotal experience~\cite{rall2016data}. Research on anthropographics suggest that, at best, there is a small effect of anthropographics on prosocial behaviors~\cite{Morais_Jansen_Andrade_Dragicevic_2021}. 
Potentially, this could be the result that data visualizations numb the viewers.

Psychology research has shown that empathy is not linear or fixed. Particular settings evoke stronger or weaker feelings of empathy. 
The identifiable victim effect is one such effect---empathy is highest for an identifiable or known person~\cite{Kogut_Ritov_2005, Vastfjall_Slovic_Mayorga_Peters_2014}. 
Even a tiny change of which grammatical articles to use can influence perception. Communicating \textit{the} (already chosen) victim is more effective than communicating that \textit{a} (will be chosen) victim~\cite{small2003helping}. If victims are perceived as one entity (e.g., a family), they will receive more donations than the same victims depicted as separate individuals~\cite{Smith_Faro_Burson_2013}. 
Empathy is highest for a single individual, and every additional individual is valued at a lesser value than the one before. Even just the move from one person in need to two people in need results in compassion fade. 
Vastfjall et al. found that donations were highest when participants were told about a single child named Rokia~\cite{Vastfjall_Slovic_Mayorga_Peters_2014}. When participants were shown both Rokia and a second child, Moussa, they donated less money than they would have donated to only Rokia. 

This trend of compassion fade continues in even larger numbers with psychic numbing, the phenomenon of people's decreasing empathy reactions to large numbers of suffering people~\cite{Slovic_2007}. The term psychic numbing was initially coined by Robert Lifton, referring to the lack of emotional response to the mass deaths due to the nuclear bombing on Hiroshima~\cite{Lifton_1982}. Generally, people show minimal differences in empathy at very large numbers of deaths. For example, feeling empathy for 200,000 lives lost is not much different from 300,000 lives lost. This could be due to a lack of capacity---humans have limited attention and imagery~\cite{Cameron_2017, Slovic_Slovic_2015}. Alternatively, this could be that people are motivated to avoid compassion for large numbers of people to avoid the emotional and (sometimes) financial cost of feeling compassion~\cite{Cameron_2017}. Psychic numbing is also observed in how people talk about tragedies in news articles and social media; larger numbers of people result in a diminished emotional response~\cite{Bhatia_Walasek_Slovic_Kunreuther_2021}. One drawback to showing the entire scope of the problem is that it can induce pseudo-inefficiency. A colossal problem makes people think that their potential action will not be useful since it is just a drop in the bucket~\cite{vastfjall2015pseudoinefficacy}. One suggested strategy to reduce psychic numbing is to reduce attention to or remove all references to the larger scope of the tragedy~\cite{Slovic_Slovic_2015}.

Statistical information has been shown to have a numbing effect on empathy. 
Participants who were shown an identifiable victim along with statistical information donated less than they would have to just an identifiable victim alone, almost decreased to the level of just presenting statistical information~\cite{Small_Loewenstein_Slovic_2007}. 
Changing emphasis on percentages compared to raw numbers can also affect audience perceptions~\cite{Fetherstonhaugh_Slovic_Johnson_Friedrich}. 
An upward trend in the data could also affect our perception of the severity of the situation, as people perceive them as more severe and donate more money than if the trend were decreasing~\cite{Erlandsson_Hohle_L2018}.


Previous literature that studies the effect of data visualizations evaluate behavior intentions or hypothetical scenarios. For example, asking how much of a donation would you allocate to two situations?
However, findings of statistical numbing use \textit{actual} donations made by participants~\cite{Small_Loewenstein_Slovic_2007}, and other research suggest that there is a difference between evaluating hypothetical donations and real donations~\cite{maier2023revisiting}. Therefore, in our study, we give participants the option of donating part of their participation incentive at the end of the experiment. The incentive is real money that the participant would receive if they chose not to donate. Our work does not rely exclusively on self-reports, which can be subject to influences of experimenter demand and social desirability.

%% file: 03-study.tex
\section{Experimental Study}
\label{section:study}

\begin{figure*}[t!]
    \centering
    \includegraphics[width=0.8\textwidth]{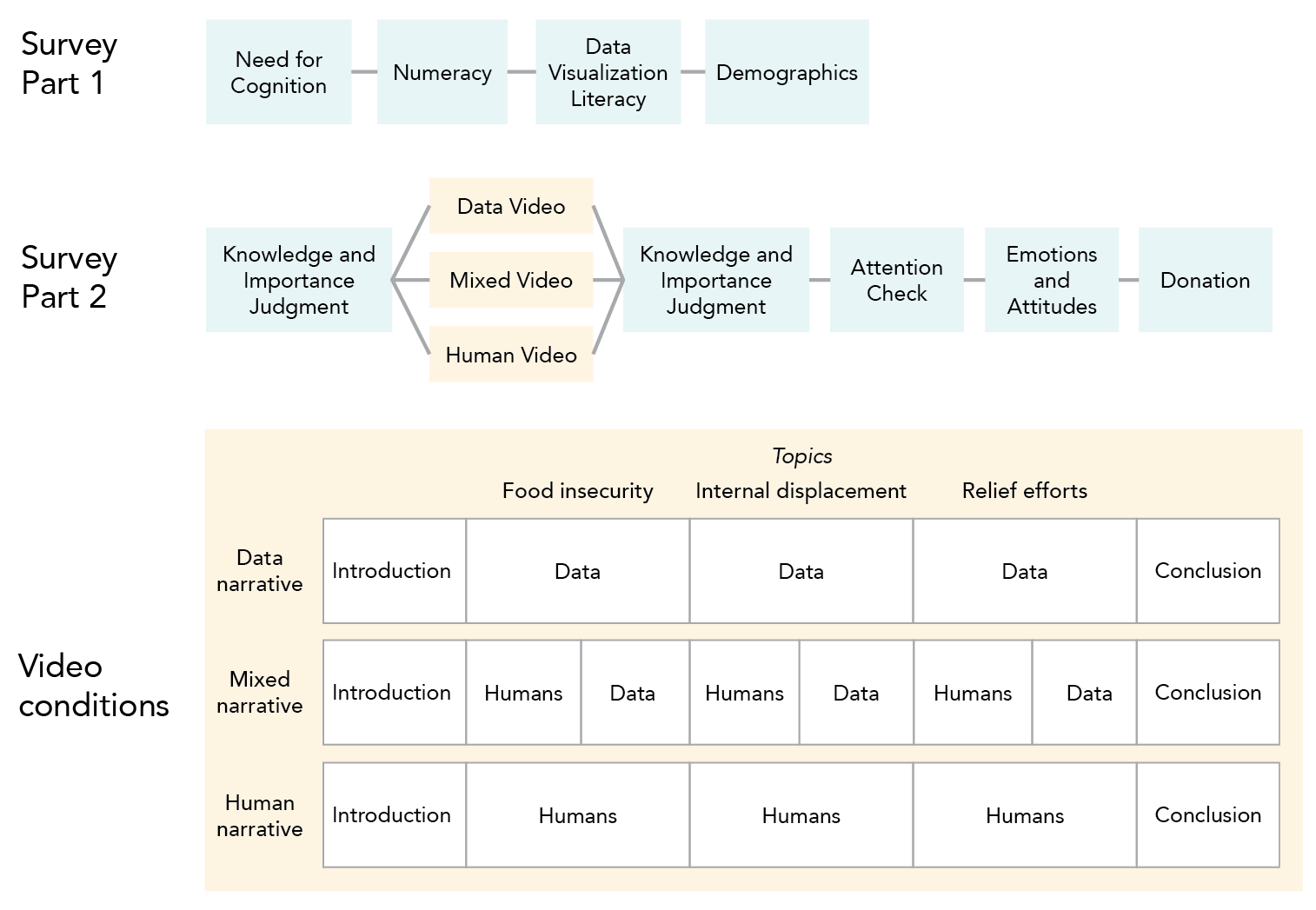}
    \caption{Experimental study design. Participants participated in a two-part survey. Participants were given one of three versions of a narrative about Somalia: data-driven narrative, human-driven narrative, or a mixed narrative of both materials.}
    \label{fig:numb-studydesign}
\end{figure*}


In Section \ref{section:related}, we discuss various ways that designers can aim to achieve affective objectives. In our study, we focus on one design in particular---the combination of data visualizations and photographs and stories about individuals. We compare this design to each type of narrative by itself, just data-driven narratives and just human-driven narratives (see Figure \ref{fig:numb-studydesign}). 
We evaluate these three designs with the goal of assessing how well they promote prosocial feelings and behaviors. Based on previous research on statistical numbing, we expect that emphasizing the \textit{humans in need} will be more effective than emphasizing \textit{the data and broader statistics}. Additionally, we are interested in exploring the combination (mixed narratives) of both of these types of information. Thus, our main research questions are: How do narrative designs of data, humans, and a combination of both influence prosocial feelings? How do they influence prosocial behaviors (donations)?
\begin{itemize}
    \item H1: Data narratives will result in weaker prosocial feelings compared to human narratives.
    \item H2: Data narratives will result in fewer donations compared to human narratives.
    \item H3: Prosocial feelings will be a mediating factor for prosocial behaviors/donations.
\end{itemize}

Under the framework of affective learning objectives \cite{lee2022affective}, we can describe our goals in the form of \textit{the viewer will [verb] [noun]}. For example, some potential affective objectives could be:
\begin{enumerate}
    \item The viewer will observe the humanitarian crisis in Somalia. 
    \item The viewer will agree that the level of poverty in Somalia is unacceptable.
    \item The viewer will believe in advocacy efforts.
    \item The viewer will act in support of advocacy by donating to the humanitarian fund. 
\end{enumerate}
In particular, we focus on the final objective listed, \textit{the viewer will donate}. Although affective objectives that aim to influence behaviors may be the most difficult to achieve, they are commonly identified as goals, either implicitly, or explicitly, as is indicated in calls to action at the end of a story. 
One benefit of separating the goals from the strategies of how we attempt to achieve them is that we can evaluate various techniques. Our goal is to disentangle the role of evoking prosocial feelings and emotions from the goal of raising donations. 

\subsection{Materials}
In this study, we measured prosocial feelings and prosocial behaviors in response to data-focused narratives and human-focused narratives.
We showed participants a video about the humanitarian crisis in Somalia of internal displacement and food insecurity resulting from a drought that affects 7 million people. 
We chose to implement our stimuli in the form of a video so that we could control how long the participant viewed each type of narrative information. Although much of the informational content on advocacy is in the form of long essays, articles, and photos, video is also a common format to produce content in. In addition, using a video would allow us to control the exposure any participant had to the study materials. With written materials, significant deviations may occur due to different reading speeds. However, this type of material might not generalize to other formats, such as interactive visuals, reports, or other designs. 

We started with an initial script for the human-driven narrative video based on a long-form humanitarian essay with stories about individual Somalis and Somali families~\cite{UNOCHA_2022}. The essay, titled ``Somalia: Hope fades as famine looms'' was created by UN OCHA, a humanitarian organization. It aimed to inform the audience about the humanitarian crisis in Somalia, as well as direct them to a donation button at the bottom of the page. This essay also contained photographs of the individuals and families that were featured in the stories. We used the essay as a starting point because it had numerous and high-quality photographs, a significant amount of written content to use, and the content was compelling and emotional. The emotional quality of the stories and the link to the donation site suggest that this would be effective content to achieve our affective goals. We use both the visual content and written content in our videos, with the written content becoming the narrative voice-over in the video while the photographs are being shown.  

In addition, we used a yearly report of the fund---the Humanitarian Response Plan~\cite{UNOCHA_2023}---as a starting point for our data-driven narrative video. The report is published each year by UN OCHA and covers provides detailed information on the crisis and data on relief efforts (111 pages in the 2023 version we used). Critically, it contains many data visualizations on the current situation in Somalia, including maps, bar charts, donut charts, bubble/area charts, icon arrays, stacked bar charts, and line charts. In addition, the Humanitarian Response Plan also included photographs of Somalis and other visual information such as a timeline of response events and organizational structure. We identified three main topics that were covered in the long-form essay and found equivalent information in the Humanitarian Response Plan. In addition, we used written content from the foreword of the report, which summarized key points about the data. This became the narrative voice-over in the video while the data visualizations were shown. We used this report as our starting point for the data condition because it was (a) authored by the same organization, UN OCHA, and (b) contained numerous and a variety of content that we could use. The data visualizations were well constructed and clearly communicated the need for humanitarian aid in Somalia. In this way, they made a logical argument about why a person should care about, and donate towards, the humanitarian crisis. In cases where we needed to supplement the existing figures in the report with additional graphics, we searched for similar data visualizations in news outlets (e.g., The Economist~\cite{EconomistSomalia}). 

Using both types of information, we constructed three versions of the video: data narrative, human narrative, and a mixed narrative combination of both. Each video has three major sections, one about internally displaced people, one about food insecurity, and one about relief efforts. Each section is approximately the same length, edited down to be 146-207 words long. All videos have the same introduction and ending, which are accompanied by photographs of Somalia, but do not focus on any individuals. The combination video contains 50\% content from each previously described video. Each video has roughly the same duration (5:39, 5:23, and 5:28). The experimental materials can be found in the supplemental materials.

\subsection{Methods}

Our experimental study is set up as a two-part survey, implemented through Prolific. We separated it into two surveys to reduce priming and fatigue effects. 
Participants were compensated with \$1.40 for each part. If participants completed both parts of the study, they were awarded a \$2.00 bonus, for a total compensation of \$4.80. Each part of the survey took about 10 minutes (part 1 mean completion time: 9.0 minutes, part 2 mean completion time: 10.6 minutes). See Figure \ref{fig:numb-studydesign} for a diagram of the survey structure. We preregistered our study design and main analyses on OSF (https://osf.io/brpf4/).

In part 1, we used established instruments to measure need for cognition~\cite{cacioppo1982need}, numeracy~\cite{schwartz1997role, lipkus2001general}, data visualization literacy~\cite{lee2016vlat, pandey2023mini}, and demographics. In particular, we were interested in measuring various individual differences that could potentially be a moderating effect on how a person might respond to a data visualization. Need for cognition measures how much a person seeks out and enjoys cognitive thinking that requires effort. As data visualization might require more cognitive effort to understand, we considered that a person's inclination towards cognitive effort would be an important factor to measure. Numeracy measures how well a person can understand numerical concepts and perform basic mathematical calculations. Data visualization literacy assesses how well a person is able to read and understand visualizations. Taking both of these together might allow us to explore if a person's ability to work with numbers and graphs would influence how they perceive the data narrative. Finally, demographics such as perceived income difficulty might be a factor in how willing a participant is to donate money. 

In part 2, we first surveyed each participant about their prior knowledge and opinion on the perceived importance of the situation in Somalia. Participants then watched one of the three video versions. Following the videos, the participant would report their knowledge and importance judgments again, as well as an attention check (they were asked for a brief textual description of what they now knew about Somalia). Because there are multiple ways to measure an emotional response, we used the same questions as in previous work that most closely matched what we wanted to extend~\cite{Small_Loewenstein_Slovic_2007, lindauer2020comparing}. In this phase, we measured prosocial feelings: upset, sympathetic, closeness, moral responsibility, and efficacy. Specifically, we used the statements:

\begin{itemize}
    \item I feel upset.
    \item I feel sympathy for people in Somalia.
    \item I feel close to people in Somalia.
    \item I have a moral responsibility to help people in Somalia. 
    \item I can make a difference in helping people in Somalia.
\end{itemize}

The response scale for these was: Not at all; Slightly; Somewhat; Moderately; Very much.

Finally, participants were given the opportunity to donate none, part or all of their \$2 bonus to Somalia. Participants were able to donate an amount between \$0.00 and \$2.00. This donation was our main measure of prosocial behavior. Importantly, we measured \textit{actual} donations. This was motivated by previous research that gave participants 5 dollars as participation compensation and then gave them the option to donate some amount of that 5 dollars to a charity~\cite{small2003helping}. In our study, we follow this design by allowing participants to donate part of their compensation to measure an actual prosocial behavior, rather than hypothetical situations or behavior intentions, which might not be influenced in the same way. This design is also motivated by past work has indicated a bias in self-reports of donations or likely donations~\cite{bekkers2011accuracy, lee2011dealing}. We note that we were limited in the \$2 bonus amount, as the Prolific system requires us to pay the participants for the survey to approve their work, but was not restrictive in whether or not we distributed the bonus. 

\subsection{Analysis}
Based on a pilot study and prior literature, we estimated that to find a small effect size of $d=0.3$ we should have a sample size of 528 participants. For the first part, we recruited 609 participants, and 546 returned to participate in the second part (10\% dropout rate). From the 546, we removed some participants from our analysis. The brief textual descriptions of what they now know about Somalia served as an attention check. Using this, we excluded 11 low-quality open-ended responses from our data analysis (e.g., sentences that do not make sense, responses obviously written by an LLM such as ChatGPT, or responses of ``I don't know anything''). In addition, 39 participants with responses more than 2 standard deviations above or below the mean completion time were removed. We also removed 36 participants who spent more than 2 standard deviations above or below the mean duration of watching the video. After the dropouts and removals, we ended with 460 valid participants, leaving us slightly under-powered (158 data narrative, 150 human narrative, 152 in the mixed narrative). 

\begin{figure}[t!]
    \centering
    \includegraphics[width=\linewidth]{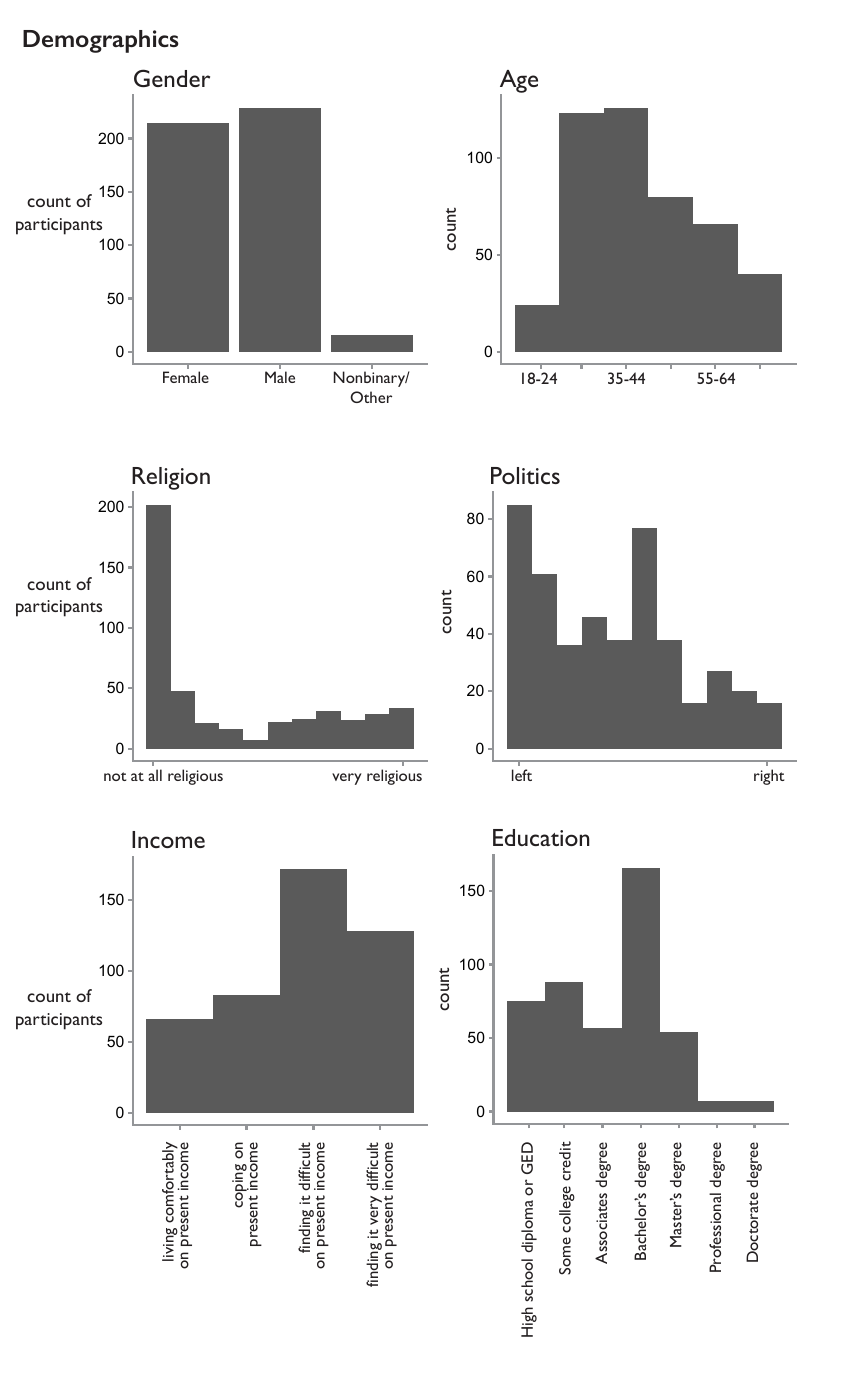}
    \caption{The overall distributions of the demographics of our participants.}
    \label{fig:demographics}
\end{figure}

Our Prolific participants were 47\% women, 50\% men, and 2\% Nonbinary/Gender Nonconforming/Genderqueer (1\% preferred not to answer). Participants had a median age of 35-44 years old. Levels of educational attainment varied from high school diploma or GED (20\%) to Doctorate degree (1\%), with the mode of Bachelor’s degree (44\%). Perception of personal income had a median rating of ``Coping on present income'' (43\%). Political affiliations had a slightly left-leaning mean of 5.0 (on a scale from 1=left to 11=right) and religiosity had a mean of 4.4, with a skewed distribution of most participants responding as not at all religious (on a scale from 1=not at all religious to 11=very religious). See Figure \ref{fig:demographics} for the full results.

Before and after the video, we asked participants to report how knowledgeable they felt about the topic and how important they perceived the situation in Somalia. Participants reported feeling more knowledgeable about Somalia after watching the video, with average ratings increasing from 1.4 to 2.3 (on a scale of 1 to 5). Similarly, judgments of importance also increased from an average of 2.8 to 3.9 (on a scale of 1 to 5). 

\begin{figure*}
    \centering
    \includegraphics[width=\textwidth]{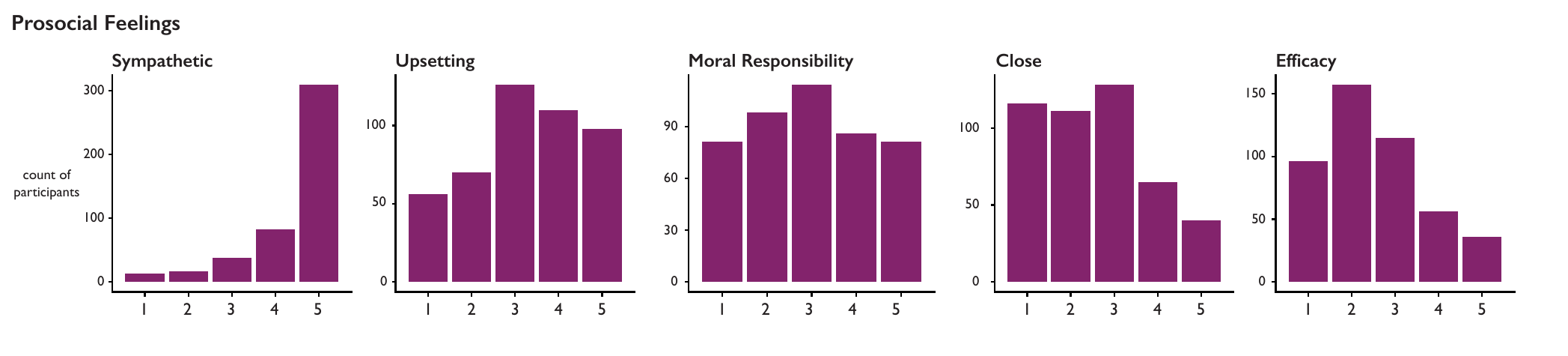}
    \caption{The overall distributions of the five prosocial feelings we measured. 1 = not at all, 5 = very much}
    \label{fig:feelings}
\end{figure*}

Participants were asked to rate five prosocial feelings (on a scale of 1 to 5). Overall, participants felt sympathetic the most (M=4.4), followed by upset (M=3.2), moral responsibility (M=2.9), closeness (M=2.6), and efficacy (M=2.5), see Figure \ref{fig:feelings}. 
We tested whether the type of narrative (data, human, or mixed) has an effect on prosocial feelings. Our hypothesis (H1) was that the human-driven narrative would evoke the strongest response for prosocial feelings. 
We ran a linear regression on each of the five prosocial feelings, and found that none of the models were statistically significant. However, the linear regression predicting \textit{upset} ratings ($R^2 = 0.0099, F(2, 457) = 2.28, p = 0.103$) showed that upset was significant ($b=0.31, p = 0.036$) but only when comparing the human-driven video ($upset M = 3.4$) to the data-driven narrative ($upset M = 3.1$). 

\begin{table}[h]
    \centering
    \begin{tabular}{cccc}
         \textbf{Variable}&  \textbf{b}&  \textbf{SE}&  \textbf{p}\\
         \hline
         (intercept)&  3.132&  0.102&  <0.001*\\
         mixed narrative&  0.107&  0.147&  0.465\\
         human narrative&  0.308&  0.146&  0.036*\\
    \end{tabular}
    \caption{Linear regression analysis summary for type of narrative predicting predicting upset ratings. The reference group is the data narrative.}
    \label{tab:regression-a}
\end{table}

\begin{figure}[b!]
    \centering
    \includegraphics[width=0.5\textwidth]{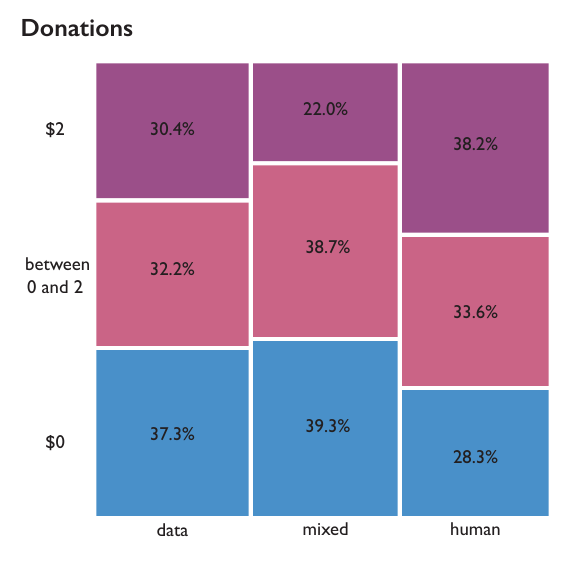}
    \caption{Donations split into groups, by all, part, or none, by type of narrative}
    \label{fig:donations}
\end{figure}

Next, we tested whether the type of video had an effect on donations (see Figure~\ref{fig:donations}). Our hypothesis (H2) was that the human-driven narrative would evoke the strongest response for prosocial behaviors.  An ANOVA found that there was a significant difference in donations between the types of narrative ($F(2, 457) = 5.53, p = 0.004$). Tukey's HSD test for pairwise comparisons showed that the significant difference was only between the mixed narrative ($M = 0.75$) and the human narrative ($M = 1.02, p = 0.003$). The data narrative ($M = 0.86$) was not significantly different from either of the other two conditions. We found that donations were not normally distributed, with 30\% of participants donating all of their bonus, 35\% donating none of their bonus, and 35\% donating part of their bonus. Therefore, we also conducted a non-parametric analysis as a sensitivity test. A Kruskal-Wallis test and a pairwise Dunn's test showed the same result: donations for the mixed narrative are lower than the human narrative ($H(2) = 9.82, p = 0.007$).

As a next step, we investigated the relationship between prosocial feelings and donations. A linear regression tested if ratings of upset, sympathy, moral responsibility, closeness, and efficacy predicted donations. 
The overall model was significant ($R^2 = 0.25, F(5, 454) = 30.9, p < 0.0001$). Feelings of upsetting ($b = 0.17, p < 0.0001$) and moral responsibility ($b = 0.20, p < 0.0001$) were both significant predictors of donations.

\subsubsection{Mediation Analysis}

Next, we explored whether the emotional response acts as a mediator variable for the relationship between the type of narrative and donation behavior (H3). One potential explanation for the influence of narrative on the audience is that the information might evoke an emotional response that then affects the behavioral response. If data and human narratives lead to different emotional responses, that could be an important factor when requesting a donation. An emotional response can mediate the attitude change and behavioral intentions~\cite{Schieferdecker_2021}. Therefore, we conducted a mediation analysis to test whether narrative conditions cause an emotional response that then drives donations.

We conducted a mediation analysis with a multicategorical independent variable (narrative type)~\cite{hayes2014statistical,lavaan} to see if feelings of upset significantly mediated the relationship between type of narrative and donations. 
Changing the video from the human narrative to the mixed narrative was associated with an $a_1=-\$0.20 (SE=0.15)$ decrease in ratings of upset. Changing the narrative from the human narrative to the data narrative was associated with an $a_2=-\$0.31 (SE=0.15)$ decrease in donations. 
An increase in ratings of upset was associated with donating $b=\$0.28 (SE=0.03)$ more money. 
The indirect effect of narrative on donations through the upset feelings pathway was not significant for the mixed narrative video $ab_1 = -0.06 (SE = 0.04)$, but was significant for the data narrative $ab_2 = -0.09 (SE = 0.04)$.
A bias-corrected bootstrapped confidence interval with 10,000 samples was above zero, 95\% CI [-0.172, -0.007]. 
Controlling for feelings of upset, the direct effect of the mixed narrative still results in significant decreases in donations, $c_1' = -\$0.26 (SE = 0.08)$, though the direct effect was not significant for the data narrative $c_2' = -\$0.10 (SE = 0.09)$.

\begin{table*}
    \centering
    \begin{tabular}{|c|c|c|c|c|c|} \hline  
         Dependent variable&  Independent variable&  parameter&  b&  se& p\\ \hline  
         upsetting&  mixed narrative&  a1&  -0.201& 0.147& 0.171\\ \hline  
         upsetting&  data narrative&  a2&  -0.308&  0.147& 0.037*\\ \hline  
         donation&  upsetting&  b&  0.284&  0.025& <0.001*\\ \hline  
         donation&  mixed narrative&  c1&  -0.257&  0.084& 0.002*\\ \hline  
         donation&  data narrative&  c2&  -0.103&  0.087& 0.237\\ \hline  
         &  &  ab1&  -0.057&  0.042& 0.172\\ \hline  
         &  &  ab2&  -0.087&  0.042& 0.039*\\ \hline  
         & & total1& -0.314& 0.094& 0.001*\\ \hline 
         & & total2& -0.190& 0.098& 0.052\\ \hline
    \end{tabular}
    \caption{Mediation analysis results summary. b estimate is unstandardized. }
    \label{tab:mediation}
\end{table*}

\begin{figure*}[t!]
    \centering
    \includegraphics[width=.75\textwidth]{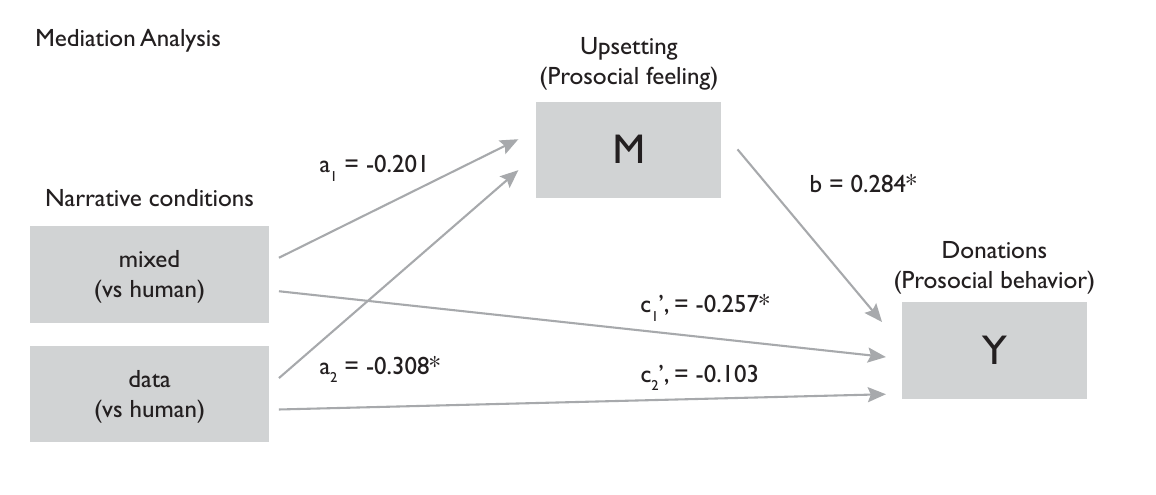}
    \caption{Estimated model coefficients from the mediation analysis. * indicates statistically significant at the p = 0.05 level. }
    \label{fig:mediation}
\end{figure*}

\subsubsection{Additional Analyses}

In addition to the main hypotheses above, we also investigated the influence of individual differences on prosocial feelings and donations. A linear regression tested if gender, age, education, income, religion, and politics predicts donations. The model was significant ($R^2 = 0.029, F(7, 435) = 2.9, p = 0.006$). We found that religion ($b = 0.03, p = 0.04$) and politics ($b = -0.05, p = 0.002$) significantly predicted donations.
Furthermore, we tested whether numeracy, data visualization literacy, or need for cognition predicted donations, but none were significant predictors of donations. 

In addition, we tested whether importance ratings and knowledge about Somalia were significant predictors of donations. Linear regression testing was used to determine if the pre-test importance, pre-test knowledge, post-test importance, and post-test knowledge predicted donations were significant ($R^2 = 21.61, F(4, 455) = 21.61, p < 0.001$). We find that post-importance rating was the only variable that predicted donations ($b=0.25, p<0.001$). This suggests that the affective impact of the video treatment is more important than the cognitive impact.

A backward stepwise linear regression was used to identify possible predictors of donations out of the following variables we measured: type of narrative, individual differences (need for cognition, data visualization literacy, numeracy), ratings of knowledge and importance (pre-test knowledge, pre-test importance, post-test knowledge, post-test importance), prosocial feelings (moral responsibility, upset, efficacy, closeness, sympathy), and demographic information (gender, age, education, religion, politics, and income). At each step, the AIC of the model was used to select the best model by iteratively removing the worst predictor variable. The final model in the stepwise linear regression was reduced to six predictors: type of narrative, moral responsibility, upsetting, post-test importance, income, and efficacy.

%% file: 04-discussion.tex
\section{Discussion}

Overall, the study results suggest that the type of narrative influences donation amounts. The human-driven narrative focused on individuals elicited the most donations. This result is not entirely surprising, as previous research in psychology demonstrating statistical numbing is what motivated this research project \cite{Small_Loewenstein_Slovic_2007}. This provides some evidence that data visualizations might numb an emotional response compared to stories and photos of individuals in need. 

We also found that prosocial feelings play an important role. Feelings of upset significantly predicted donation amounts. Other prosocial feelings that influence donations are moral responsibility and perceptions of importance. The analyses provide evidence that feelings of upset are also the highest for the human-driven narrative. This provides some evidence that the human-driven narrative is the most effective at influencing prosocial feelings and prosocial behaviors, though the type of narrative did not have any effect on the other feelings.

Surprisingly, we found that the mixed narrative resulted in the lowest average donation. This was unexpected, because we initially conceptualized the mixed narrative as falling between the data-driven and the human-driven narrative. One way to consider the types of narratives is to create a continuous variable of how human-focused the video is: 0\% (data narrative), 50\% (mixed narrative), or 100\%. However, the results suggest that these videos are categorically different rather than ordinal. The mixed narrative resulted in the lowest donations, the human narrative had the highest donations, and the data narrative was in the middle. In previous research, the combination condition of multiple appeals was between both conditions (individuals and statistics~\cite{Small_Loewenstein_Slovic_2007}, emotional and rational appeals~\cite{lindauer2020comparing}. However, we found that our combination condition performed the worst.

One hypothesis for why this might be is that multiple messages are are more complex and are likely to be less effective than one single message. In Made to Stick, messages are described as most effective if they are simple~\cite{heath2007made}. In our stimuli, it is possible that the mixed narrative was perceived as multiple messages, whereas the data narrative and the human narrative were more cohesive. That said, we did intentionally aim to control for the topic subject matter, and had three topics that were the same across all videos. We intended the mixed narrative to present the same message (``there is high food insecurity'') but with two types of information (data visualizations, photographs and stories) supporting that message.  
Another potential reason why the mixed narrative performed the worst is that the transitions from people to data happened several times. Because we had three topics, the combination video alternated from people to data several times throughout the video. This may have led to a different experience compared to the other videos. Additional work may help identify whether this phenomenon can be countered by alternative design strategies. However, the result does indicate that designs that combined representations may end up being worst-of-both-worlds rather than the best.

\subsection{Data Visualization Objectives}

In this study, our focus was prioritizing and evaluating affective objectives over cognitive objectives. Our goal was to investigate the prosocial feelings and behaviors in response to a humanitarian crisis. Our affective goals could be phrased as learning objectives, such as, The viewer will \textit{observe the humanitarian crisis}, \textit{agree that the level of poverty in Somalia is unacceptable}, and \textit{donate money to support humanitarian relief}. By separating out the goals from the techniques, we can consider the role of each more deeply. We conceptualized that evoking emotion is one strategy that could be used to achieve these goals. Additionally, communicating the magnitude of the crisis could be another strategy to achieve these goals. We evaluate each strategy against our affective goals. 

This study focuses on the affective outcomes of the narratives. However, humanitarian organizations might have goals to achieve cognitive outcomes, such as recalling key information, comparing categories, or analyzing the information. While the data visualizations communicated numerical information, such as the percentage of Somalis facing food insecurity or the geographic location of internal displacement, this was not the focus of our study. In our study, we did not evaluate cognitive outcomes. Future work could also measure the effectiveness of these cognitive outcomes, by asking participants to recall key facts or predict future trends. There may be a trade-off to achieve both affective and cognitive goals. For example, there could be varying ways to construct a video that has a combination of data visualizations and stories of individuals. Our study used a 50/50 split of both of these types of information, but future work could explore 20/80 splits, or a more creative way to communicate information through audio and visual channels.   

Future work should explore the goals of data visualization designers in more detail. It is likely that data visualization designers have both affective and cognitive goals, as well as other aims. For example, designers could have an objective to be perceived as data-driven, trustworthy, and credible. Using data visualizations would be effective for some of these goals. In addition, designers could have goals related to increasing engagement, views, likes, and shares of their work. Future work should explore how designers prioritize and rank goals when they have more than one. 

\subsection{Limitations and Future Directions}

In our study, we chose to replicate UN OCHA data visualizations and supplement them with those from news outlets. While we included a range of commonly used data visualizations, such as bar charts, maps, and line charts, these results might not generalize to other data visualization designs. Additionally, these more standard and traditional data visualization designs that we used do not take advantage of many of the anthropographic~\cite{Morais_Jansen_Andrade_Dragicevic_2020} or affective~\cite{lan2023affective} design potentials. While the data visualizations still communicate the logical argument, or the logos rhetoric, they might not combine well with the content about the individuals as less-emotive visuals. 
There are many expressive and effective data visualizations in the wild. Future work can look into novel designs that include other factors that might influence affective outcomes, such as interactive design, scrollytelling, novel encodings, and more~\cite{lan2023affective}. We presented the data-driven narratives and the human-driven narrative segments in our combination video sequentially, but there is an opportunity to integrate the data and stories about people more closely. For example, Periscopic's U.S. Gun Deaths identifies individuals in the data as they are represented in the visualization~\cite{Periscopic_2013}. In this way, the information of individuals is integrated into the data visualization, which may lead to a better user experience.

In our study, we attempted to measure various factors of individual differences that could have affected a person's openness to information in the form of data-driven narratives and human-driven narratives. We hypothesized that the need for cognition, data visualization literacy, and numeracy could affect the way people receive data visualization information. However, none of these factors appeared to have an effect on our models. Future directions can consider additional factors of individual differences to see if there is another factor that might mediate the response people have to different types of media. 

Designers may intuitively want to combine data visualizations, photographs, and stories of individuals together. However, there is no comprehensive evaluation of the benefits and drawbacks of this strategy. Data visualization designers continue to create visualizations about humanitarian crises, even though prior psychology research shows that statistical numbing may reduce prosocial feelings and behaviors. This paper supports this claim that data visualizations may not be the most effective technique to achieve affective objectives for increasing prosocial behaviors. However, future work should aim to disentangle the effect of multiple aspects of this domain, such as various visualization techniques, designer goals, and individual differences within the audience.

%% file: 05-conclusion.tex
\section{Conclusion}

Although psychology research has shown that statistics numb prosocial feelings and behaviors, data visualization practitioners continue to use graphs and charts to show statistics on humanitarian issues. If designers have affective goals to increase empathy and prosocial behaviors, data visualizations may not be the most effective technique to achieve those goals. We explore the pitfalls and potentials of data visualizations to evoke prosocial feelings and increase prosocial behaviors. Our work provides some evidence that stories and photographs of individuals in need continues to be the most effective way to elicit prosocial behaviors such as donations.

%% file: bib.bib
@book{bogre2012photography,
  title={Photography as activism: images for social change},
  author={Bogre, Michelle},
  year={2012},
  publisher={Routledge}
}

@article{cacioppo1982need,
  title={The need for cognition.},
  author={Cacioppo, John T and Petty, Richard E},
  journal={Journal of personality and social psychology},
  volume={42},
  number={1},
  pages={116},
  year={1982},
  publisher={American Psychological Association}
}

@misc{Demir_2017, 
    title={Alan Kurdi}, 
    rights={Creative Commons Attribution-ShareAlike License}, 
    url={https://en.wikipedia.org/w/index.php?title=File:Alan_Kurdi_lifeless_body.jpg&oldid=813344027}, 
    note={Page Version ID: 813344027}, 
    author={Demir, Nilüfer}, 
    year={2017}, 
    month=dec, 
    language={en} 
}

@article{barry2013framing,
  title={Framing childhood obesity: How individualizing the problem affects public support for prevention},
  author={Barry, Colleen L and Brescoll, Victoria L and Gollust, Sarah E},
  journal={Political Psychology},
  volume={34},
  number={3},
  pages={327--349},
  year={2013},
  publisher={Wiley Online Library}
}

@article{garreton2023attitudinal,
  title={Attitudinal effects of data visualizations and illustrations in data stories},
  author={Garret{\'o}n, Manuela and Morini, Francesca and Celhay, Pablo and D{\"o}rk, Marian and Parra, Denis},
  journal={IEEE Transactions on Visualization and Computer Graphics},
  year={2023},
  publisher={IEEE}
}

@article{small2003helping,
  title={Helping a victim or helping the victim: Altruism and identifiability},
  author={Small, Deborah A and Loewenstein, George},
  journal={Journal of Risk and uncertainty},
  volume={26},
  pages={5--16},
  year={2003},
  publisher={Springer}
}

@article{vastfjall2015pseudoinefficacy,
  title={Pseudoinefficacy: negative feelings from children who cannot be helped reduce warm glow for children who can be helped},
  author={V{\"a}stfj{\"a}ll, Daniel and Slovic, Paul and Mayorga, Marcus},
  journal={Frontiers in psychology},
  volume={6},
  pages={134125},
  year={2015},
  publisher={Frontiers}
}

@article{maier2023revisiting,
  title={Revisiting and Rethinking the Identifiable Victim Effect: Replication and Extension of Small, Loewenstein, and Slovic (2007)},
  author={Maier, Maximilian and Wong, Yik Chun and Feldman, Gilad},
  journal={Collabra: Psychology},
  volume={9},
  number={1},
  year={2023},
  publisher={University of California Press}
}

@misc{Merite_2021, 
 title={Climate Change Knows No Borders}, 
 howpublished={\url{https://www.instagram.com/p/CQ_uSGCh66N/}}, 
 author={Merite, Gabrielle}, 
 year={2021}, 
 month={Jul} 
}

@misc{Voila2023,
    howpublished={https://chezvoila.com/project/freedom-denied/}, 
    title={Freeing the US from its culture of detention}, 
    journal={Voilà:}, 
    author={Voilà},
    language={en-CA} 
}

@book{siegler2022freedom,
  title={Freedom Denied: How the Culture of Detention Created a Federal Jailing Crisis. },
  author={Siegler, Alison},
  year={2022},
  publisher={University of Chicago Law School Federal Criminal Justice Clinic},
  url={http://freedomdenied.law.uchicago.edu/}
}

@incollection{ganesh2015communicating,
  title={Communicating Gender: The Challenges of Visualizing Information for Advocacy},
  author={Ganesh, Maya Indira and Sobliye, Gabi},
  booktitle={Diversity and Design},
  pages={137--151},
  year={2015},
  publisher={Routledge}
}

@article{rall2016data,
  title={Data visualization for human rights advocacy},
  author={Rall, Katharina and Satterthwaite, Margaret L and Pandey, Anshul Vikram and Emerson, John and Boy, Jeremy and Nov, Oded and Bertini, Enrico},
  journal={Journal of Human Rights Practice},
  volume={8},
  number={2},
  pages={171--197},
  year={2016},
  publisher={Oxford University Press}
}

@article{lee2022affective,
  title={Affective learning objectives for communicative visualizations},
  author={Lee-Robbins, Elsie and Adar, Eytan},
  journal={IEEE Transactions on Visualization and Computer Graphics},
  volume={29},
  number={1},
  pages={1--11},
  year={2022},
  publisher={IEEE}
}

@ARTICLE{adar2021,
  author={Adar, Eytan and Lee, Elsie},
  journal={IEEE Transactions on Visualization and Computer Graphics}, 
  title={Communicative Visualizations as a Learning Problem}, 
  year={2021},
  volume={27},
  number={2},
  pages={946-956},
  doi={10.1109/TVCG.2020.3030375}
}

@article{lan2023affective,
  title={Affective Visualization Design: Leveraging the Emotional Impact of Data},
  author={Lan, Xingyu and Wu, Yanqiu and Cao, Nan},
  journal={IEEE Transactions on Visualization and Computer Graphics},
  year={2023},
  publisher={IEEE}
}

@article{lindauer2020comparing,
  title={Comparing the effect of rational and emotional appeals on donation behavior},
  author={Lindauer, Matthew and Mayorga, Marcus and Greene, Joshua and Slovic, Paul and V{\"a}stfj{\"a}ll, Daniel and Singer, Peter},
  journal={Judgment and Decision making},
  volume={15},
  number={3},
  pages={413--420},
  year={2020},
  publisher={Cambridge University Press}
}

@article{pandey2023mini,
  title={Mini-{VLAT}: A Short and Effective Measure of Visualization Literacy},
  author={Pandey, Saugat and Ottley, Alvitta},
  journal={arXiv preprint arXiv:2304.07905},
  year={2023}
}

@article{lee2016vlat,
  title={{VLAT}: Development of a visualization literacy assessment test},
  author={Lee, Sukwon and Kim, Sung-Hee and Kwon, Bum Chul},
  journal={IEEE transactions on visualization and computer graphics},
  volume={23},
  number={1},
  pages={551--560},
  year={2016},
  publisher={IEEE}
}

@article{lipkus2001general,
  title={General performance on a numeracy scale among highly educated samples},
  author={Lipkus, Isaac M and Samsa, Greg and Rimer, Barbara K},
  journal={Medical decision making},
  volume={21},
  number={1},
  pages={37--44},
  year={2001},
  publisher={Sage Publications Sage CA: Thousand Oaks, CA}
}

@article{schwartz1997role,
  title={The role of numeracy in understanding the benefit of screening mammography},
  author={Schwartz, Lisa M and Woloshin, Steven and Black, William C and Welch, H Gilbert},
  journal={Annals of internal medicine},
  volume={127},
  number={11},
  pages={966--972},
  year={1997},
  publisher={American College of Physicians}
}

@article{hayes2014statistical,
  title={Statistical mediation analysis with a multicategorical independent variable},
  author={Hayes, Andrew F and Preacher, Kristopher J},
  journal={British journal of mathematical and statistical psychology},
  volume={67},
  number={3},
  pages={451--470},
  year={2014},
  publisher={Wiley Online Library}
}

@Article{lavaan,
    title = {{lavaan}: An {R} Package for Structural Equation
      Modeling},
    author = {Yves Rosseel},
    journal = {Journal of Statistical Software},
    year = {2012},
    volume = {48},
    number = {2},
    pages = {1--36},
    url = {http://www.jstatsoft.org/v48/i02/},
}

@misc{UNOCHA_2022, 
    title={Somalia: Hope fades as famine looms}, 
    url={https://unocha.exposure.co/somalia-hope-fades-as-famine-looms/photos/8095661}, 
    journal={Exposure}, 
    author={{United Nations Office for the Coordination of Humanitarian Affairs (UN OCHA)}}, 
    year={2022}, 
    month=nov, 
    language={en} 
}

@article{bekkers2011accuracy,
  title={Accuracy of self-reports on donations to charitable organizations},
  author={Bekkers, Ren{\'e} and Wiepking, Pamala},
  journal={Quality \& Quantity},
  volume={45},
  pages={1369--1383},
  year={2011},
  publisher={Springer}
}

@article{lee2011dealing,
  title={Dealing with social desirability bias: An application to charitable giving},
  author={Lee, Zoe and Sargeant, Adrian},
  journal={European Journal of Marketing},
  volume={45},
  number={5},
  pages={703--719},
  year={2011},
  publisher={Emerald Group Publishing Limited}
}

@misc{UNOCHA_2023, 
    title={Humanitarian Response Plan Somalia}, 
    author={{United Nations Office for the Coordination of Humanitarian Affairs (UN OCHA)}}, 
    year={2023}, 
    month=feb 
}

@article{EconomistSomalia, 
    ISSN={0013-0613},  
    url={https://www.economist.com/graphic-detail/2022/12/19/somalia-is-on-the-brink-of-famine}, 
    journal={The Economist}, 
    author={Graphic Daily},
    title={Somalia is on the brink of famine},
    year={2022}, 
    month=dec 
}

@book{heath2007made,
  title={Made to stick: Why some ideas survive and others die},
  author={Heath, Chip and Heath, Dan},
  year={2007},
  publisher={Random House}
}

@book{Cameron_2017, 
    title={Compassion Collapse}, 
    volume={1}, 
    url={http://oxfordhandbooks.com/view/10.1093/oxfordhb/9780190464684.001.0001/oxfordhb-9780190464684-e-20}, 
    DOI={10.1093/oxfordhb/9780190464684.013.20}, 
    publisher={Oxford University Press}, 
    author={Cameron, C. Daryl}, 
    year={2017}, 
    month={Oct} 
}

@misc{NYT_names_2020, 
    title={Remembering the 100,000 Lives Lost to Coronavirus in America}, 
    url={https://www.nytimes.com/interactive/2020/05/24/us/us-coronavirus-deaths-100000.html}, 
    journal={New York Times}, 
    author={{New York Times}}, 
    year={2020}, 
    month={May} 
}

@misc{Lewis_2020, 
    title={Where Are the Photos of People Dying of Covid?}, 
    url={https://www.nytimes.com/2020/05/01/opinion/coronavirus-photography.html}, 
    journal={New York Times}, 
    author={Lewis, Sarah}, 
    year={2020}, 
    month={May} 
}

@inbook{Alamalhodaei_Alberda_Feigenbaum_2020, 
    place={NL Amsterdam}, 
    title={Humanizing data through ‘data comics’: An introduction to graphic medicine and graphic social science}, 
    ISBN={978-94-6372-290-2}, 
    url={https://www.aup.nl/en/book/9789463722902}, 
    DOI={10.5117/9789463722902}, 
    booktitle={Data Visualization in Society}, 
    publisher={Amsterdam University Press}, 
    author={Alamalhodaei, Aria and Alberda, Alexandra and Feigenbaum, Anna}, 
    year={2020}, 
    month={Apr}, 
    pages={347–365} 
}

@inproceedings{Bach_Wang_Farinella_Murray-Rust_Riche_2018, 
    place={Montreal QC Canada}, 
    title={Design Patterns for Data Comics}, 
    ISBN={978-1-4503-5620-6}, 
    url={https://dl.acm.org/doi/10.1145/3173574.3173612}, 
    DOI={10.1145/3173574.3173612},
    booktitle={Proceedings of the 2018 CHI Conference on Human Factors in Computing Systems}, 
    publisher={ACM}, 
    author={Bach, Benjamin and Wang, Zezhong and Farinella, Matteo and Murray-Rust, Dave and Henry Riche, Nathalie}, 
    year={2018}, 
    month={Apr}, 
    pages={1–12} 
}

@article{Black_2019, 
    title={Black Girl Magic}, 
    url={https://www.instagram.com/blackbythenumbers/}, 
    journal={Instagram}, 
    author={{Black by the Numbers}}, 
    year={2019}, 
    month={Dec} 
}

@misc{Kristof_2009, 
    title={Nicholas Kristof’s Advice for Saving the World}, 
    url={https://www.outsideonline.com/outdoor-adventure/nicholas-kristofs-advice-saving-world/}, 
    journal={Outside Online}, 
    author={Kristof, Nicholas}, 
    year={2009}, 
    month={Nov} 
}

@article{Grubbs_Madrid_Yahnke_Lin_2013, 
    title={Out of Sight, Out of Mind}, 
    url={http://drones.pitchinteractive.com}, 
    journal={Pitch Interactive}, 
    author={Grubbs, Wesley and Madrid, Katarina and Yahnke, Nick and Lin, Pei-Yu}, 
    year={2013} 
}

@article{kostelnick2016,
  title={The re-emergence of emotional appeals in interactive data visualization},
  author={Kostelnick, Charles},
  journal={Technical Communication},
  volume={63},
  number={2},
  pages={116--135},
  year={2016},
  publisher={Society for Technical Communication}
}

@unpublished{Rost2017,
 title= {A Data Point Walks Into a Bar: Designing Data For Empathy},
 author = {Rost, Lisa},
 year = {2017},
 note= {OpenVis Conference},
 URL= {https://www.youtube.com/watch?v=8XgF-RmNwUc},
}

@book{Tactile_Technology_Collective_2013, 
    place={Bangalore}, 
    title={Visualizing Information for Advocacy}, 
    author={{Tactile~Technology~Collective}}, 
    year={2013} 
}

@misc{Zeeberg_2016, 
    title={How Images Trigger Empathy}, 
    url={https://www.theatlantic.com/science/archive/2016/01/cultivate-empathy-photograph/422793/}, 
    journal={The Atlantic}, 
    author={Zeeberg, Amos}, 
    year={2016}, 
    month={Jan} 
}

@article{Bhatia_Walasek_Slovic_Kunreuther_2021, title={The More Who Die, the Less We Care: Evidence from Natural Language Analysis of Online News Articles and Social Media Posts}, volume={41}, ISSN={0272-4332, 1539-6924}, DOI={10.1111/risa.13582}, number={1}, journal={Risk Analysis}, author={Bhatia, Sudeep and Walasek, Lukasz and Slovic, Paul and Kunreuther, Howard}, year={2021}, month={Jan}, pages={179–203} }

@article{Boy_Pandey_Emerson_2017, title={Showing People Behind Data: Does Anthropomorphizing Visualizations Elicit More Empathy for Human Rights Data?}, author={Boy, Jeremy and Pandey, Anshul Vikram and Emerson, John}, journal={Proceedings of the 2017 CHI Conference on Human Factors in Computing Systems}, year={2017}, pages={13} }

@phdthesis{campbell_2018, place={Boston, Massachusettes}, title={Feeling Numbers}, school={Northeastern U.}, author={Campbell, Sarah}, year={2018} }

@article{Campbell_Offenhuber_2019, title={Feeling numbers: The emotional impact of proximity techniques in visualization}, volume={25}, ISSN={0142-5471, 1569-979X}, DOI={10.1075/idj.25.1.06cam}, number={1}, journal={Information Design Journal}, author={Campbell, Sarah and Offenhuber, Dietmar}, year={2019}, month={Dec}, pages={71–86} }

@misc{Slobin, title={What If the Data Visualization Is Actually People?}, url={https://source.opennews.org/articles/what-if-data-visualization-actually-people/}, abstractNote={Sarah Slobin discovers that all the facts and numbers didn’t add up to the humans in her story}, author={Slobin, Sarah}, year={2014}}

@inbook{Emerson_Satterthwaite_Pandey_2018, edition={1}, title={The Challenging Power of Data Visualization for Human Rights Advocacy}, ISBN={978-1-316-83895-2}, url={https://www.cambridge.org/core/product/identifier/978131683895223CN-bp-8/type/book_part}, DOI={10.1017/9781316838952.008}, booktitle={New Technologies for Human Rights Law and Practice}, publisher={Cambridge University Press}, author={Emerson, John and Satterthwaite, Margaret L. and Pandey, Anshul Vikram}, year={2018}, month={Apr}, pages={162–187} }

@article{Erlandsson_Hohle_L2018, title={The rise and fall of scary numbers: The effect of perceived trends on future estimates, severity ratings, and help-allocations in a cancer context}, volume={48}, ISSN={00219029}, DOI={10.1111/jasp.12552}, number={11}, journal={Journal of Applied Social Psychology}, author={Erlandsson, Arvid and Hohle, Sigrid Møyner and Løhre, Erik and Västfjäll, Daniel}, year={2018}, month={Nov}, pages={618–633} }

@article{Fetherstonhaugh_Slovic_Johnson_Friedrich, title={Insensitivity to the Value of Human Life: A Study of Psychophysical Numbing}, journal={Journal of Risk and Uncertainty}, author={Fetherstonhaugh, David and Slovic, Paul and Johnson, Stephen M and Friedrich, James}, pages={18}, year={1997} }

@misc{Harris2015, title={Connecting with the Dots}, url={https://source.opennews.org/articles/connecting-dots/}, author={Harris, Jacob}, year={2015}, journal={Source} }

@article{Hullman_Diakopoulos_2011, title={Visualization Rhetoric: Framing Effects in Narrative Visualization}, volume={17}, ISSN={1077-2626}, DOI={10.1109/TVCG.2011.255}, number={12}, journal={IEEE Transactions on Visualization and Computer Graphics}, author={Hullman, J. and Diakopoulos, N.}, year={2011}, month={Dec}, pages={2231–2240} }

@article{Kogut_Ritov_2005, title={The “identified victim” effect: an identified group, or just a single individual?}, volume={18}, ISSN={0894-3257, 1099-0771}, DOI={10.1002/bdm.492}, number={3}, journal={Journal of Behavioral Decision Making}, author={Kogut, Tehila and Ritov, Ilana}, year={2005}, month={Jul}, pages={157–167} }

@misc{Kontinentalist_2020, title={Abandoned at sea: The desperate journeys of Rohingya refugees}, url={https://kontinentalist.com/stories/the-rohingya-in-myanmar-a-refugee-crisis-at-sea}, author={Kontinentalist}, year={2020}, month={Dec} }

@article{Liem_Perin_Wood_2020, title={Structure, Empathy, and Attitude}, journal={Eurographics Conference on Visualization (EuroVis) 2020}, author={Liem, J and Perin, C and Wood, J}, year={2020}, pages={13} }

@article{Morais_Dandara_Sousa_Andrade_2020, title={Evaluating a Situated and Physical Anthropographic: An In-the-Wild Study}, url={http://rgdoi.net/10.13140/RG.2.2.10719.69283}, DOI={10.13140/RG.2.2.10719.69283},  journal={Proceedings of the 2019 CHI Conference on Human Factors in Computing Systems}, author={Luiz Morais and Dandara Sousa and Andrade, Nazareno}, year={2020} }

@article{Maier_Slovic_Mayorga_2017, title={Reader reaction to news of mass suffering: Assessing the influence of story form and emotional response}, volume={18}, ISSN={1464-8849, 1741-3001}, DOI={10.1177/1464884916663597}, number={8}, journal={Journalism}, author={Maier, Scott R and Slovic, Paul and Mayorga, Marcus}, year={2017}, month={Sep}, pages={1011–1029} }

@article{Morais_Jansen_Andrade_Dragicevic_2020, title={Showing Data about People: A Design Space of Anthropographics}, ISSN={1077-2626, 1941-0506, 2160-9306}, DOI={10.1109/TVCG.2020.3023013}, journal={IEEE Transactions on Visualization and Computer Graphics}, author={Morais, Luiz and Jansen, Yvonne and Andrade, Nazareno and Dragicevic, Pierre}, year={2020}, pages={1–1} }

@article{Morais_Jansen_Andrade_Dragicevic_2021, title={Can Anthropographics Promote Prosociality? A Review and Large-Sample Study}, author={Morais, Luiz and Jansen, Yvonne and Andrade, Nazareno and Dragicevic, Pierre}, year={2021}, pages={19}, journal={Proceedings of the 2021 CHI Conference on Human Factors in Computing Systems} }

@article{Pandey_Manivannan_Nov_Satterthwaite_Bertini_2014, title={The Persuasive Power of Data Visualization}, volume={20}, ISSN={1077-2626}, DOI={10.1109/TVCG.2014.2346419}, number={12}, journal={IEEE Transactions on Visualization and Computer Graphics}, author={Pandey, Anshul Vikram and Manivannan, Anjali and Nov, Oded and Satterthwaite, Margaret and Bertini, Enrico}, year={2014}, month={Dec}, pages={2211–2220} }

@misc{Periscopic_2013, title={U.S. Gun Deaths}, url={https://guns.periscopic.com/}, journal={Periscopic}, author={Periscopic}, year={2013} }

@article{Schieferdecker_2021, title={Passivity in the face of distant others’ suffering: an integrated model to explain behavioral (non-)response}, volume={45}, ISSN={2380-8985, 2380-8977}, DOI={10.1080/23808985.2021.1908837}, number={1}, journal={Annals of the International Communication Association}, author={Schieferdecker, David}, year={2021}, month={Jan}, pages={20–38} }

@misc{Slovic_2007, title={Psychic Numbing and Genocide: (718332007-003)}, url={http://doi.apa.org/get-pe-doi.cfm?doi=10.1037/e718332007-003}, DOI={10.1037/e718332007-003}, publisher={American Psychological Association}, author={Slovic, Paul}, year={2007} }

@article{Smith_Faro_Burson_2013, title={More for the Many: The Influence of Entitativity on Charitable Giving}, volume={39}, ISSN={0093-5301, 1537-5277}, DOI={10.1086/666470}, number={5}, journal={Journal of Consumer Research}, author={Smith, Robert W. and Faro, David and Burson, Katherine A.}, year={2013}, month={Feb}, pages={961–976} }

@article{Small_Loewenstein_Slovic_2007, title={Sympathy and callousness: The impact of deliberative thought on donations to identifiable and statistical victims}, volume={102}, ISSN={07495978}, DOI={10.1016/j.obhdp.2006.01.005},  number={2}, journal={Organizational Behavior and Human Decision Processes}, author={Small, Deborah A. and Loewenstein, George and Slovic, Paul}, year={2007}, month={Mar}, pages={143–153} }

@article{Vastfjall_Slovic_Mayorga_Peters_2014, title={Compassion Fade: Affect and Charity Are Greatest for a Single Child in Need}, volume={9}, ISSN={1932-6203}, DOI={10.1371/journal.pone.0100115}, number={6}, journal={PLoS ONE}, author={Västfjäll, Daniel and Slovic, Paul and Mayorga, Marcus and Peters, Ellen}, editor={Lamm, Claus}, year={2014}, month={Jun}, pages={e100115} }

@book{Slovic_Slovic_2015, place={Corvallis, OR}, title={Numbers and nerves: information, emotion, and meaning in a world of data}, ISBN={978-0-87071-776-5}, author={Slovic and Slovic}, publisher={Oregon State University Press}, year={2015} }

@article{Lifton_1982, title={Beyond psychic numbing: A call to awareness}, volume={52}, number={4}, journal={American Journal of Orthopsychiatry}, author={Lifton, Robert}, year={1982}, pages={619–629} }
